# Diffuse-Charge Effects on the Transient Response of Electrochemical Cells

M. van Soestbergen[1], P. M. Biesheuvel[2] and M. Z. Bazant[3]

[1]Materials innovation institute, Mekelweg 2, 2628 CD Delft, the Netherlands and Department of Precision and Microsystems Engineering, Delft University of Technology, Mekelweg 2, 2628 CD Delft, the Netherlands.

[2]Department of Environmental Technology, Wageningen University, Bomenweg 2, 6703 HD Wageningen, the Netherlands. [3]Departments of Chemical Engineering and Mathematics, Massachusetts Institute of Technology, Cambridge, MA 02139, USA.

**Abstract**

We present theoretical models for the time-dependent voltage of an electrochemical cell in response to a current step, including effects of diffuse charge (or "space charge") near the electrodes on Faradaic reaction kinetics. The full model is based on the classical Poisson-Nernst-Planck equations with generalized Frumkin-Butler-Volmer boundary conditions to describe electron-transfer reactions across the Stern monolayer at the electrode surface. In practical situations, diffuse charge is confined to thin diffuse layers (DLs), which poses numerical difficulties for the full model but allows simplification by asymptotic analysis. For a thin quasi-equilibrium DL, we derive effective boundary conditions on the quasi-neutral bulk electrolyte at the diffusion time-scale, valid up to the transition time, where the bulk concentration vanishes due to diffusion limitation. We integrate the thin DL problem analytically to obtain a set of algebraic equations, whose (numerical) solution compares favorably to the full model. In the Gouy-Chapman and Helmholtz limits, where the Stern layer is thin or thick compared to the DL, respectively, we derive simple analytical formulae for the cell voltage versus time. The full model also describes the fast initial capacitive charging of the DLs and super-limiting currents beyond the transition time, where the DL expands to a transient non-equilibrium structure. We extend the well-known Sand equation for the transition time to include all values of the super-limiting current beyond the diffusion-limiting current.

## 1 Introduction

Time-dependent models for electrochemical cells are widely used in science and technology. In the field of power sources, the charge/discharge cycle of batteries[1-5] and the startup behavior of fuel cells[6] are important topics. Time-dependent models with electrochemical reactions have been used to describe e.g. light-emitting devices,[7] metal deposition in nanotubes,[8] ion intercalation in nanoparticles,[9,10] corrosion,[11] ion exchange membranes,[12] and electrokinetic micro-pumps.[13,14]

Sand's classical theory[15] of the transient voltage of a flat electrode in response to a current step is based on a similarity solution of the diffusion equation in a semi-infinite, one-dimensional domain with a constant-flux boundary condition.[16] Experimental data from chronopotentiometry of electrochemical cells[16] and galvanostatic intermittent titration of rechargeable batteries[2] is routinely fitted to Sand's formula, but discrepancies can arise because the theory assumes linear response in a neutral bulk solution and ignores diffuse charge near the electrodes. Charge relaxation in a thin diffuse layer (DL) can be included empirically by placing a capacitor in parallel with the Faradaic current[16,17] or more systematically, by analyzing the underlying transport equations, which has been extensively developed for blocking, ideally polarizable electrodes.[18-22] For non-blocking electrodes passing

Faradaic current, these approaches must be extended to include the Frumkin correction and other nonlinear modifications of the reaction rate associated with diffuse charge.[23]

The usual starting point to describe the transport of ions is the Poisson-Nernst-Planck (PNP) theory, which leads to a set of coupled, nonlinear differential equations. These equations are difficult to solve numerically due to the formation of local space-charge regions near the electrodes, typically in very thin DLs whose width is on the order of the Debye screening length (~1-100 nm). Within the DLs, gradients in concentration and potential are very steep, compared to gradual variations in the quasi-neutral bulk electrolyte at the scale of the feature size of the system, often orders of magnitude larger. In numerical models this would lead to the requirement of a very fine spacing of grid points near the electrodes, especially in two and three dimensions.[24] To circumvent this problem a vast amount of previous work assumes electro-neutrality throughout the complete electrolyte phase, thus neglecting the DLs.[3-6,16,25,26] The DLs, however, cannot be neglected as they influence the charge transfer rate at the electrodes and make significant contribution to the cell voltage.[23,27-34]

To describe Faradaic reaction rates at the electrodes, it is reasonable to assume that the charge transfer occurs at some atomic distance away from the electrodes, at the "reaction plane" (for an atomically flat electrode), which serves as the edge of the continuum region representing the electrolyte (Fig. 1). The reaction plane is commonly equated with the "outer Helmholtz plane" or "Stern plane" of closest approach of solvated ions to the surface, so the electrolyte region is separated from the metal electrode by a thin Stern monolayer of solvent molecules on the electrode.[35] The Stern layer is often viewed as uncharged and polarizable, with a reduced dielectric constant compared to the bulk, due to dipole alignment in the large normal field of the DL. This model, which we adopt below, neglects specific adsorption of ions, which break free of solvation and adsorb onto the surface at the "inner Helmholtz plane" within the Stern layer, either as an intermediate step in Faradaic reactions or without any charge transfer. In that case, the electrolyte is effectively separated from the electrode by a thin dielectric coating, the "Stern layer", and we postulate that electron-transfer reactions occur across this layer, biased by the local voltage drop.[27] More sophisticated models of the DL are available, including various models with finite-sized ions and solvent molecules, as reviewed in Ref. 14. Even the simplest approach used here, however, provides a rich microscopic description of Faradaic reactions at the electrode surface that goes well beyond the standard approach in electrochemistry of applying the Butler-Volmer equation across the entire electrochemical double layer,[16] i.e. across both the DL and Stern layer (without the "Frumkin correction"[23,27]).

Regardless of the detailed model of the Stern layer, its local potential drop, which biases electron tunneling and Faradaic reaction rates, has a complicated nonlinear dependence on the accumulated charge and voltage across the DL. In order to complete the model, an electrostatic boundary condition is required at the Stern plane, which we take to be a Robin-type condition equating the normal electric field with the Stern-layer voltage, effectively assuming a constant field in the Stern layer.[23,28,30,31] Combining this boundary condition self-consistently with Butler-Volmer kinetics for the charge-transfer reaction, biased by the Stern voltage, leads to a unified microscopic model, which we refer to as the "generalized Frumkin-Butler-Volmer" (gFBV) equation.[23,27-34] (See Ref 23 for a historical review.) In our work, the gFBV equation is beneficial over the traditionally used Butler-Volmer equation since it provides a natural boundary condition for the full PNP theory, including diffuse charge.[23,30,31]

In the current work we develop a simple semi-analytical theory of the time-dependent response of an electrochemical cell with thin DLs (compared to geometrical features of the cell) and validate it against numerical solutions of the full PNP-gFBV equations. The simple model circumvents the numerical problems in solving the full model by using matched asymptotic expansions to integrate across the DL to derive effective boundary conditions on the quasi-neutral bulk solution. We focus on the typical situation where the neutral salt concentration just outside the thin DL remains non-negligible (prior to diffusion limitation), in which case the diffuse charge profile of the DL remains in quasi-equilibrium,[21-23,30-34,36-42] even while passing a significant (but below limiting) current. Time-dependent breakdown of quasi-equilibrium and concomitant expansion of the DL has just begun to be analyzed for blocking electrodes,[18,21] but this is a major complication for electrochemical cells passing Faradaic current, which to our knowledge has only been analyzed in steady-state situations.[32]

The thin-DL model extends classical time-dependent models based on the hypothesis of electro-neutrality throughout the complete electrolyte phase, by treating reaction kinetics at the electrodes in a self-consistent way, properly accounting for diffuse charge (i.e., accounting for the DLs). The effective equations and boundary conditions of the thin-DL model are very general and can be applied to any transient problem, but we illustrate their use in the standard case of a suddenly applied, constant current between parallel-plate electrodes. Unlike transient problems involving a suddenly applied voltage across parallel plates[18,19,20] or around a metallic particle or microstructure,[39,40,43] which involve nontrivial, time-dependent coupling of the DL and quasi-neutral bulk regions, the situation of one-dimensional transient conduction at constant current offers a well known analytical simplification, namely, the salt concentration of the quasi-neutral bulk evolves according to the diffusion equation (Fick's law) with constant flux boundary conditions due to mass transfer at the electrodes,[18,44-50] while the time evolution of the DLs (and thus the total voltage of the cell) is slaved to that of the bulk concentration. The bulk diffusion problem has an exact solution in terms of an infinite series,[46-49] which describes the spreading and collision of diffusion layers from the two electrodes. Prior to collision of the diffusion layers (and prior to the transition time discussed below), the solution can also be approximated more accurately by similarity solutions for semi-infinite domains near each electrode.[14,15] These classical solutions are used to infer the mass transfer properties of an electrolyte from experimental transient measurements.[51-53]

In our analysis, we neglect the very early stage of transient response, where the double layers charge capacitively, prior to the onset of bulk diffusion. In this phase of the dynamics, the bulk concentration remains nearly uniform, and thus the bulk acts like a resistor in series with the double layer capacitors. The associated "RC" time scale can be expressed as $\lambda_D L / D$, where $D$ is the electrolyte diffusivity, $L$ is the electrode spacing, and $\lambda_D$ is the Debye screening length, which sets the thickness of the diffuse double layers[18]. At low voltages or in the absence of Faradaic reactions, this is the relevant time-scale for transient response, e.g. in high-frequency impedance spectroscopy experiments[36,54,55] or induced-charge electrokinetics, where high-frequency alternating currents are applied.[13,38-40] The bulk diffusion time-scale $L^2/D$, which is much larger than the RC time for thin DLs (by a factor of $L/\lambda_D$), becomes important in presence of Faradaic reactions[23,30-33] or at large applied voltages for blocking electrodes where the DL adsorbs neutral salt such that the bulk region becomes depleted.[18,21,22,38] In this work we will focus on the diffusion time-scale, as we are interested in the start-up behavior of electrochemical cells at longer times than the relaxation time, all the way up to the steady-state. Consequently, the transient solutions we present here can be regarded as an extension into the time-dependent domain of the steady-state solutions reported previously in refs. 23,31 and 32.

To show the accuracy of the simplified approach for thin DLs we will compare results with full model predictions, and show that they compare very favorably for all parameter settings that we investigate, except for very short time-scales where the relaxation time-scale applies. Furthermore, we will show results for the classical transition time, or "Sand's time", when the salt concentration approaches zero at either electrode due to an applied current exceeding diffusion limitation.[16,46,56] For such large currents, we show that the cell potential reaches an infinite value for the thin-DL limit, while it remains finite for any non-zero DL thickness, due to the expansion of the DL into a different non-equilibrium structure. Analogous effects, first studied in the context of over-limiting direct currents in electrodialysis,[57] have recently been analyzed for steady-state problems involving Faradaic reactions[31,32] and for time-dependent problems involving large AC voltages with blocking electrodes,[21] but we are not aware of any prior modeling of transient response to over-limiting current in electrochemical cells with Faradaic reactions.

## 2    Theory

In this section we will discuss a time-dependent model for a planar one-dimensional electrochemical cell containing a binary electrolyte, i.e. the electrolyte contains only two species, namely, cations, at a concentration $C_c(X,t)$, and anions, at concentration $C_a(X,t)$. The continuum electrolyte phase is bounded by the two reaction planes (one on each electrode), where we assume that the only reaction is the formation (at the anode) or removal (at the cathode) of the cations (Fig. 1). The anion is assumed to be inert and as a result the total number of anions in the system is conserved; no such number constraint applies for the reactive cations. This situation describes, for example, an electrolytic cell, such as Cu|CuSO$_4$|Cu transporting Cu$^+$ with metal deposition at the cathode and dissolution at the anode, or a galvanic cell, such as a PEM fuel cell conducting protons across a membrane, or a thin-film Li-ion battery shuttling Li$^+$ between intercalation electrodes. The charge free Stern layer is located in between the reaction planes and the electrodes, and is treated mathematically via the boundary conditions.

In the full Poisson-Nernst-Planck (PNP) model the electrolyte phase is modeled by the same set of equations in the entire electrolyte phase, all the way up to the reaction planes, irrespective of the amount of local charge separation (which is low in the bulk and high near the reaction planes). However, in the simplified model that we will discuss, the continuum electrolyte phase is divided in two domains. The first domain is the outer region where electro-neutrality can be assumed, a region which we will also refer to as the bulk region. The second domain is the inner region (found on both electrodes) where we have a non-zero space-charge density, such that it captures the DLs. We will first present the full model for ion transport and the electrochemical reactions based on the full PNP-gFBV theory, from which we derive the simplified, semi-analytical, model using the singular perturbation theory. For both models we will focus on the situation where a constant current is prescribed and we will follow the development of the cell potential as function of time.

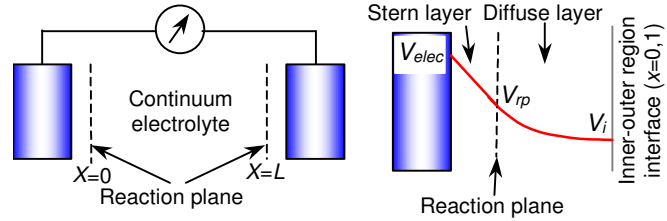

*Fig. 1. (Color online) Schematics of the electrochemical cell. In the simplified model the electrolyte phase is split in an outer (bulk) and inner (diffuse layer) region, the latter present on both electrodes. The potential in the electrode ($V_{elec}$) decays linearly across the Stern layer to the reaction plane ($V_{rp}$) and decays further across the diffuse layer to the inner-outer region interface ($V_i$).*

## 2.1 PNP-gFBV theory

To describe ion transport in the electrolyte phase, the Nernst-Planck (NP) equation is the standard model, in which it is assumed that ions are ideal point charges. (For a review of more general, modified NP equations for concentrated electrolytes with finite-sized ions, see Ref. 14.) In this work, we also neglect convection fluxes of ions, due to flow of the solvent. In this case the NP-equation combined with a mass balance, leads to

$$\dot{C}_i = -\partial_X J_i = \partial_X \cdot [D_i(\partial_X C_i + z_i \cdot C_i \cdot f \cdot \partial_X V)] \tag{1}$$

where $\dot{C}_i$ is the time-derivative of the ion concentration $C_i$, $J_i$ is the ion flux, $D_i$ the ion diffusion coefficient, $z_i$ the valence of the ions, $V$ the local electrostatic potential in Volts, $f$ equals $e/k_B T$, with $e$ the elementary charge, $k_B$ Boltzmann's constant and $T$ temperature. Subscript $i$ either denotes the reactive cations, i.e., $c$, or the inert anions, i.e., $a$. In Eq. (1), $X$ is the spatial coordinate which runs from the first reaction plane at $X=0$ to the second reaction plane at $X=L$ (Fig. 1) and thus applies up to the plane of closest approach for both the cat- and anions, irrespective of whether the ion is inert or electrochemically reactive. In the local-density mean-field approximation, the mean electrostatic potential is related to the charge density by Poisson's equation

$$\varepsilon \cdot \partial_X^2 V = -e \cdot \sum_i z_i C_i \tag{2}$$

where $\varepsilon$ is the dielectric permittivity. If we combine eqs. (1) and (2) we obtain the full PNP theory.

The flux of cations at the reaction planes due to the electrochemical reaction of these ions to a metal atom (or other reductant species) in the electrode ($Me$) can be described by the gFBV equation[23,27-34]

$$J_F = K_R C_O \exp(-\alpha_R \cdot f \cdot \Delta V_S) - K_O C_R \exp(\alpha_O \cdot f \cdot \Delta V_S) \tag{3}$$

where $\Delta V_S$ is the potential drop across the Stern layer ($\Delta V_S = V_{elec} - V_{rp}$, see Fig. 1), $K_i$ are kinetic constants and $\alpha_i$ are the transfer coefficients (here $\alpha_i = \frac{1}{2}$ is assumed[31]), subscript O and R denote the oxidized state (or oxidation reaction) and reduced state (reduction reaction), respectively. In the present work we consider that the ion is converted to a neutral metal atom and incorporated in the electrode (and vice-versa), i.e. $C_c + e^- \leftrightarrow Me$. Therefore we can assume a constant metal atom concentration (reduced state), so from this point onward we can make the replacement $J_O = C_R K_O$ for a constant oxidation reaction rate.

Next we discuss the dimensionless parameters that are applicable to eqs. (1)-(3) and which we will use in the remainder of this work, following Refs. 18,21,23,30-32,39 and 41. First we have the dimensionless electrostatic potential, $\phi = f \cdot V$, the average total concentration of ions, $c = (C_c + C_a)/(2 \cdot C_\infty)$, and diffuse charge density, $\rho = (C_c - C_a)/(2 \cdot C_\infty)$, where $C_\infty$ is the initial average ion concentration, set by the ionic strength of the bulk electrolyte. Note that the average ion concentration is only constant for the anions, but not necessarily for the cations, due to the absence of a constraint on equal formation and removal rates, which can lead to nonzero total charge in the electrolyte (balanced by opposite charges on the electrodes).[23,30-32] These two rates become equal only when the steady state is reached, while in the time up to the steady-state an excess or deficit of reactive ions is produced. Further, we introduce the length scale $x = X/L$. At this length scale, the Debye length, $\lambda_D = \sqrt{\varepsilon k_B T / 2e^2 C_\infty}$, becomes $\epsilon = \lambda_D/L$. Next, we scale the Stern layer thickness to the Debye length, such that its dimensionless equivalent becomes $\delta = \lambda_s/\lambda_D$. We scale time to the diffusion time scale discussed above, so the dimensionless time is $\tau = t \cdot D/L^2$. Note that here we implicitly assume that the diffusion coefficient of both ionic species are equal and are independent of the ion concentration. Finally, we scale the ion flux to the diffusion-limiting current, $j = J_i/J_{lim} = J_i \cdot L/4DC_\infty$, such that the dimensionless kinetic constants of the gFBV equation become $k_R = K_R C_\infty/J_{lim}$ and $j_O = J_O/J_{lim}$.

Making these substitutions, the PNP equations take the dimensionless form:

$$\dot{c} = \partial_x(\partial_x c + \rho \cdot \partial_x \phi) \tag{4a}$$

$$\dot{\rho} = \partial_x(\partial_x \rho + c \cdot \partial_x \phi) \tag{4b}$$

$$\epsilon^2 \partial_x^2 \phi = -\rho \tag{4c}$$

The boundary conditions that apply at the reaction planes for the transport equations are given by

$$\partial_x c + \rho \cdot \partial_x \phi = -2 j_F \tag{5a}$$

$$\partial_x \rho + c \cdot \partial_x \phi = -2 j_F \tag{5b}$$

where the Faradaic rate of formation (or removal) of cations at the reaction planes is given by

$$\pm j_F = k_R(c + \rho) \cdot \exp(-\tfrac{1}{2}\Delta\phi_S) - j_O \cdot \exp(\tfrac{1}{2}\Delta\phi_S), \tag{6}$$

where the $\pm$–sign refers to the positive value at position $x$=1 and the negative value at position $x$=0. Note that we define a flux of cations from $x$=0 to $x$=1 as positive.

The potential drop across the Stern layer equals the potential in the adjacent metal phase, $\phi_{elec}$, minus the potential at the reaction plane, $\phi_{rp}$, and relates to the electrical field strength at the reaction plane according to[23,30-32]

$$\Delta \phi_S = \pm \epsilon \cdot \partial_x \phi, \tag{7}$$

where the ±–sign again refers to the positive value at position $x=1$ and the negative value at position $x=0$. The boundary conditions for the Poisson equation can be determined as follows, namely by making use of the fact that in the electrolyte the imposed electrical current $i$ is equal to the sum of the ionic conduction current and the Maxwell displacement current.[11,30] This equality is generally valid at each time and each position. At the electrodes, the ion conduction current is also always exactly equal to the Faradaic (i.e., electrochemical or charge-transfer) reaction rate, $j_F$. The conduction current in the electrolyte results from the migration of ions, with initially both the cat- and anions contributing to this current, while at the steady-state only the flux of cations remains. The Maxwell current (which, dimensionally, is $-\varepsilon \cdot \partial_x \dot{V}$) originates from changes with respect to time of the electrical field strength, and thus vanishes at steady-state. As a result we have[11,12,30,36]

$$i = -\tfrac{1}{2}(\partial_x \rho + c \cdot \partial_x \phi) - \tfrac{1}{2} \cdot \epsilon^2 \frac{d}{d\tau} \partial_x \phi \tag{8}$$

for the current in the electrolyte phase, which becomes

$$i = j_F - \tfrac{1}{2} \cdot \epsilon^2 \frac{d}{d\tau} \partial_x \phi, \tag{9}$$

at the reaction planes. Instead of using eq. (9) directly, we integrate and use[11,12]

$$\partial_x \phi = \frac{2}{\epsilon^2} \int_0^\tau [\pm j_F(\tau') - i] d\tau', \tag{10}$$

for the electrical field strength, $\partial_x \phi$, at the reaction planes. We will refer to the set of equations described above as the full PNP model.

### 2.2 Thin-DL limit

Next, we use singular perturbation theory (matched asymptotic expansions) to simplify the full PNP model by integrating across the DLs to obtain a set of equations and effective boundary conditions for the quasi-neutral bulk region, which is more tractable for numerical or analytical solution. In the case of transient one-dimensional response to a constant current between parallel plates, we will see that the thin-DL model can be solved analytically, at least in implicit algebraic form. This well known approach provides a systematic mathematical basis for the physical intuition that the problem can be solved in two distinct domains, namely the "inner region" of the DLs and "outer region" of the quasi-neutral bulk solution (Fig. 1), and appropriately matched to construct uniformly valid approximations, in the asymptotic limit $\epsilon \to 0$ of thin DLs compared to the system size.[18,30-33,37,58]

We first discuss the inner solution, describing the structure of the DLs, which are present at both electrodes. The inner regions are defined by the coordinate system $y=x/\epsilon$ for the region near $x=0$ and $y=(1-x)/\epsilon$ for the region near $x=1$. Conversion of eq. (9) to inner coordinates results in the fact that the Maxwell current will vanish for $\epsilon \to 0$ (thin-DL limit). Consequently, the Faradaic current becomes equal to the current imposed on the system, i.e. $j_F = \pm i$. Then we can eliminate c+ρ at the reaction plane from eq. (6) as described next. Namely, we substitute eq. (4)c in eq. (4)a, convert to the inner coordinates and perform the integration with the matching conditions $\partial_y \phi = \partial_y^2 \phi = 0$ for $y \to \infty$, which yields

$$c = \tfrac{1}{2}(\partial_y \phi)^2 + c_i, \tag{11}$$

for the variation across the inner region of the concentration $c$ (note again, only for the condition that $\epsilon \to 0$), where $c_i$ denotes the concentration at the inner-outer region interface. Next, we substitute eq. (4)c into eq. (4)b, rewrite and convert to the inner region coordinates, then substitute eq. (11) and finally perform the integration with the matching conditions, $\partial_y \phi = \partial_y^3 \phi = 0$ for $y \to \infty$, to obtain

$$\partial_y^3 \phi - \tfrac{1}{2}(\partial_y \phi)^3 - c_i \cdot \partial_y \phi = 0, \tag{12}$$

for the thin-DL limit. Eq. (12) is identical to the steady-state Smyrl-Newman master equation[30,32,33,37] converted to the inner coordinate system for $\epsilon \to 0$, cf. Bonnefont et al.[30] eq. (33). Consequently, as for the steady-state, it is possible to derive from Eq. (12) the classical approximation that the DLs are in quasi-equilibrium (with a Boltzmann distribution for a dilute solution) in the asymptotic limit $\epsilon \to 0$,[30-33,37,45] even in the presence of a non-zero normal current, until the nearby bulk salt concentration reaches very small $O(\epsilon^{2/3})$ values. We will not show this derivation here as it is carried out in detail by Bonnefont et al.[30] At larger currents (beyond the transition time discussed below), the double layer loses its quasi-equilibrium structure,[37] and matched asymptotic expansions for over-limiting currents at Faradaic electrodes[32] or ion-exchange membranes[58] have revealed a more complicated nested boundary-layer structure for the DLs, including an intermediate, non-equilibrium extended space charge layer.

In this work, we only consider the asymptotic limit of thin quasi-equilibrium DLs, such that an infinite voltage would be required to surpass the diffusion-limited current and force the DL out of equilibrium. In this common situation, we can relate $c+\rho$ at the inner-outer region interface to $c+\rho$ at the reaction plane using the Boltzmann distribution for ions as ideal point charges,[23,30,34]

$$c_{rp} + \rho_{rp} = c_i \cdot \exp(-\Delta\phi_{DL}), \tag{13}$$

where subscript $rp$ refers to the reaction plane and $\Delta\phi_{DL}=\phi_{rp}-\phi_i$, which is the potential drop across the DL from the reaction plane, $\phi_{rp}$, to the inner-outer region interface, $\phi_i$ (Fig. 1). Note that in eq. (13) we implicitly assume $\rho=0$ in the outer region as eq. (4)c vanishes there for $\epsilon \to 0$. Next, we can use the Gouy-Chapman theory to determine the potential drop across the Stern layer according to[23,30,34]

$$\Delta\phi_S = 2\sqrt{c_i}\delta \sinh(\tfrac{1}{2}\Delta\phi_{DL}), \tag{14}$$

which is again valid when the DLs are in quasi-equilibrium and follows from Eq. (12) as described in ref. 30. Note that the dimensionless Stern layer thickness $\delta$ depends on the ion concentration via the Debye screening length $\lambda_D \propto 1/\sqrt{c_i}$. In eq. (14) we use $\sqrt{c_i}$ to correct $\delta$ for any variations in the Debye length due to variations of the ion concentration at the inner-outer region interface.[23] Although eqs. (13) and (14) are equilibrium properties of the DL, they have also proven to be very useful in describing the steady-state current of electrochemical cells,[23,30-33] for the reasons giving above. For thin DLs, there is no direct effect of the Faradaic current on the concentration profiles near the electrodes at leading order in the asymptotic limit $\epsilon \to 0$ when considering the inner region coordinate system.[30-33] Nevertheless, the magnitude of the potential drop across the Stern layer, and thus across the DL, is indirectly influenced by the Faradaic current via the gFBV equation and Stern boundary condition. Next, we can substitute eqs. (13) and (14) into eq. (6), which concludes the mathematical description of the two inner regions. Finally, we require the concentration, $c_i$, at the inner-outer region interface (matching condition), which we will explain next.

To determine the concentration at the inner-outer interface we require the outer region solution. For $\epsilon \to 0$ it follows from eq. (4)c that the space-charge density vanishes throughout the complete outer region. As a result the derivates of the space-charge with respect to time, $\dot{\rho}$, and the spatial coordinate, $\partial_x \rho$, will vanish as well. Consequently, eq. (4)b will vanish and eq. (4)a results in the linear diffusion equation, $\dot{c} = \partial_x^2 c$,[41,42,44-50] which is mathematically equivalent to Fick's second law for the diffusion of neutral particles.[48,59] The same diffusion equation also applies here to the bulk salt concentration, $c = (C_c + C_a)/(2 \cdot C_\infty)$, at leading order, thus recovering the classical model for a neutral electrolyte.[16,45,59] (For unequal diffusivities in a neutral binary electrolyte, the associated bulk salt diffusion coefficient is the "ambipolar diffusivity".[45]) The corresponding boundary conditions are given by eq. (5)a, which also simplify due to the vanishing space-charge density in the outer region for $\epsilon \to 0$, namely they become, $\partial_x c = -2 j_F$. The ability to replace the full PNP equations in the bulk of the cell by Fick's law for diffusion is an important simplification from the thin-DL limit, since the number of unknown fields is reduced from three to one in the outer region. Namely, we solve for the concentration $c$, which is the unknown field variable, while $\rho$ equals zero and the potential drop across the outer region can be determined from an algebraic equation as discussed below. Furthermore, we can make use of various exact solutions to the diffusion equation in one-dimension for various types of boundary conditions.

In this work, we apply a constant current to a cell with planar electrodes, as explained above. Note that for $\epsilon \to 0$ the second term in eq. (9), i.e. the Maxwell term, vanishes, such that $j_F = \pm i$. With these assumptions, the diffusion equation has an exact solution in terms of an infinite series[48,49] (which can be systematically derived, e.g. by Finite Fourier Transform[59]):

$$c(x,\tau) = 1 + 2 \cdot i \cdot \left\{ \frac{1}{2} - x - \sum_{n=1}^{n=\infty}(f_n \cdot \cos(2 \cdot N \cdot x)) \right\}, \quad (15)$$

where $f_n = \exp(-4 \cdot N^2 \cdot \tau)/N^2$ and $N = \frac{1}{2}\pi \cdot (2n-1)$, a solution which has been applied to model the transient current in electrochemical cells for at least a century on the basis of electro-neutrality throughout the complete electrolyte phase.[46,47,50] Here we apply eq. (15) in a model which considers space charge regions and Stern layers as well.

Due to the applied current we have either an injection or removal of salt at the edges of the outer region resulting in gradual variations of the salt concentration, which are initially localized near the edges of the outer region in the so-called diffusion layer. Eq. (15) describes the spreading and collision of two diffusion layers from the electrodes and it can be truncated at a small number of terms in the long-time limit ($\tau \gg 1$). Eventually, the exponential term in eq. (15) vanishes when $\tau \to \infty$, leaving $c(x) = 1 + i \cdot (1 - 2x)$, which is exactly the classical steady-state solution for planar electrochemical cells.[3,23,30-33] Note that the steady-state solution is physically valid (with positive concentration at the cathode) only below the diffusion-limited current, i.e. $i$<1. Larger transient currents are possible, but lead to vanishing salt concentration at the "transition time" discussed below.

For early times, prior to collision of the diffusion layers, the series in eq. (15) converges very slowly, and it is more accurate to use similarity solutions for semi-infinite diffusion layers. Close to each electrode, the bulk concentration is well approximated by the classical solution to the diffusion equation in a semi-infinite domain with a constant-flux boundary condition.[14,15] In order to construct a uniformly valid approximation for early times, we add two such similarity solutions to the initial constant concentration, one at the anode for an enrichment layer and one at the cathode for a depletion layer, resulting in

$$c(x,\tau) = 1 + 4 \cdot i \cdot \sqrt{\tau} \cdot \left\{ \text{ierfc}\left(\frac{x}{2 \cdot \sqrt{\tau}}\right) - \text{ierfc}\left(\frac{1-x}{2 \cdot \sqrt{\tau}}\right) \right\}, \qquad (16)$$

where $\text{ierfc}(x) = 1/\sqrt{\pi} \cdot \exp(-x^2) - x \cdot \text{erfc}(x)$. (See Fig. 5b below.) We focus on the early time regime below when analyzing the transition time and proceed to use the series solution for more typical situations, below the limiting current.

We can now use the set of eqs. (6),(13)-(15) under the condition that $j_F = \pm i$ to solve for the potential drop across the DL and Stern layer, as function of time and imposed current. The potential drop across the outer region for $\epsilon \to 0$ follows from eq. (8) by integration, namely,

$$\Delta\phi_{\text{outer}} = \int_{x=0}^{x=1} \frac{2 \cdot i}{c(x,\tau)} dx, \qquad (17)$$

where $c(x,\tau)$ follows from either eq. (15) or (15). In deriving eq. (17) we make use of the vanishing space-charge density, $\rho$, as mentioned previously. Furthermore, in eq. (17) the term $2/c(x,\tau)$ can be regarded as a local Ohmic resistance for the outer region. For the steady-state we can use $c(x) = 1 + i \cdot (1 - 2x)$ in eq. (17) and obtain the classical result $\Delta\phi_{\text{outer}} = 2 \cdot \tanh^{-1}(i)$.[3,23,30-33] Finally, we can define the potential drop across the complete system, i.e. the cell potential, as

$$\phi_{\text{cell}} = [\Delta\phi_S + \Delta\phi_{\text{DL}}]_{x=0} + \Delta\phi_{\text{outer}} - [\Delta\phi_S + \Delta\phi_{\text{DL}}]_{x=1}. \qquad (18)$$

We can now solve eqs. (6), (13)-(15), (17) and (18) as a small set of non-linear algebraic equations to describe the dynamic problem in case of the limit of $\epsilon \to 0$, which we will refer to as the thin-DL limit.

### 2.3 GC- and H-limit

Analytical solutions for the voltage against current curve for the steady-state have been presented in literature for two limits based on the Stern boundary condition, namely, the Gouy-Chapman (GC) limit $\delta \to 0$ and the Helmholtz (H) limit $\delta \to \infty$.[23,31] In the GC-limit, the reaction plane coincides with the electrode surface, and the Stern layer does not sustain any voltage drop. For this limit we can derive the potential drop across the DL directly from eq. (6), substituting eq. (13) and $\Delta\phi_S$=0, such that we obtain,

$$\Delta\phi_{DL,GC} = \ln\left(\frac{k_{R,m} c_{i,m}}{j_{O,m} \mp i}\right), \tag{19}$$

where $c_{i,m}$ is the concentration at the inner-outer region interface and subscript $m$ either denotes the anode (A) at $x=0$ or the cathode (C) at $x=1$. In the GC-limit the effect of a non-zero space-charge density in the DL is at maximum due to the absence of the potential drop across the Stern layer. In contrast, in the opposite H-limit this effect completely vanishes as the potential drop in this case is completely across the Stern layer, i.e. we assume an infinite Stern layer thickness relative to the thickness of the DL. Consequently, the H-limit coincides with models based on electro-neutrality, where zero space-charge density is assumed for the complete electrolyte phase.[23] The potential drop across the Stern layer in the H-limit can be derived from eq. (6) if we assume that the reaction plane coincides with the inner-outer region interface, such that,

$$\Delta\phi_{S,H} = 2\ln\left(\frac{\pm i + \sqrt{i^2 + 4 j_{O,m} k_{R,m} c_{i,m}}}{2 j_{O,m}}\right). \tag{20}$$

The concentration $c_{i,m}$ can be determined by either substituting $x=0$ or $x=1$ into eq. (15), which yields

$$c_i(\tau) = \pm i \cdot \left\{1 - 2 \cdot \sum_{n=1}^{n=\infty} f_n(\tau)\right\} + 1, \tag{21}$$

where the $\pm$ sign refers to $x=0$ and $x=1$, respectively. From $\tau \sim 0.15$ onward eq. (21) can be approximated by the first term of the summation only,[46] such that

$$c_i(\tau) \approx 1 \pm g(\tau) \cdot i, \tag{22}$$

where $g(\tau) = 1 - \frac{8}{\pi^2} \exp(-\pi^2 \cdot \tau)$. Note when $\tau \to \infty$ (thus $g(\tau) \to 1$) we obtain the steady-state solution, $c_i \approx 1 \pm i$.[23,30-33]

To obtain the potential drop across the complete cell we finally need to calculate the potential drop across the outer region. However, substitution of eq. (15) into eq. (17) does not result in an analytic solution, not even for $n=1$. Therefore we approximate the distribution of ions across the outer region as a linear function according to

$$c(x,\tau) = c(0,\tau) + \{c(1,\tau) - c(0,\tau)\} \cdot x, \tag{23}$$

which after substitution of eq. (22) yields

$$c(x,\tau) \approx i \cdot g(\tau) \cdot (1 - 2x) + 1. \tag{24}$$

Finally, we can substitute eq. (24) into eq. (17) to obtain an analytical approximation for the potential drop across the outer (bulk) region

$$\Delta\phi_{outer} \approx \frac{1}{g(\tau)} \ln\left(\frac{1 + g(\tau) \cdot i}{1 - g(\tau) \cdot i}\right) = \frac{2}{g(\tau)} \tanh^{-1}[g(\tau) \cdot i]. \tag{25}$$

We can now combine eq. (18) and eq. (25) with either eqs. (19) or (20) to obtain

$$\phi_{cell,GC} = \phi_0 + \ln\left(\frac{1 + i/j_{O,C}}{1 - i/j_{O,A}}\right) + 2\frac{1 + g(\tau)}{g(\tau)} \tanh^{-1}[[g(\tau) \cdot i] \tag{26}$$

for the cell potential in case of the GC-limit and,

$$\phi_{cell,H} = \phi_0 + 2\sinh^{-1}\left(\frac{i}{\sqrt{\beta_A(1+g(\tau)\cdot i)}}\right) + 2\sinh^{-1}\left(\frac{i}{\sqrt{\beta_C(1-g(\tau)\cdot i)}}\right) + 2\frac{1+g(\tau)}{g(\tau)}\tanh^{-1}[g(\tau)\cdot i] \quad (27)$$

for the H-limit, where $\beta_m = 4j_{O,m}k_{R,m}$ and $\phi_0 = \ln(j_{O,C}k_{R,A}/j_{O,A}k_{R,C})$, which is the open cell potential. For $\tau \to \infty$ we have $g(\tau) \to 1$, such that eqs. (26) and (27) coincide with their steady-state equivalents, i.e. eqs. (35) and (36) of ref. 23 except for a sign reversal of all terms due to the reversed definition of the cell potential in ref. 23.

We can see from eq. (26) that its second term becomes negligible when $j_{O,m} \gg i$, i.e. if the oxidation kinetic constants are high. The second and third term of eq. (27) become negligible when $\beta_m \gg i^2/(1 \pm g(\tau) \cdot i)$. Consequently, for high kinetic rate constants the value of the cell potential as predicted by the GC- and H-limits coincide.[31] As a result, the effect of the DL and Stern layer will vanish for high reaction rate kinetics. In the opposite regime, when the kinetic rates are very low, the GC-limit will show a reaction limiting current if $j_{O,m} < |i|$,[23,31] a limitation that is absent in the H-limit.

## 2.4 Transition time

Next we discuss the concept of a transition time (refs. 16 p. 307 and 46 p. 207). At the transition time the concentration, $c$, reaches zero at the inner-outer region interface at position $x$=0 for negative current or at $x$=1 for positive current. When the concentration, $c$, becomes zero the potential across the outer region according to eq. (17) will diverge such that the solution for the cell potential will show an asymptotic behavior at the transition time. We can obtain a relation to determine the transition time by substituting zero concentration at $x$=1 for a positive current (or equivalently at $x$=0 for a negative current) into eq. (15), which results in

$$\sum_{n=1}^{n=\infty} \frac{\exp\{-\pi^2(2n-1)^2\tau_{tr}\}}{(2n-1)^2} = \frac{\pi^2}{8}\left(1 - \frac{1}{|i|}\right), \quad (28)$$

where the transition time, $\tau_{tr}$, is an implicit function of the current, $i$.

We will now derive two approximate but explicit solutions for the transition time. In the first order approximation of eq. (28) ($n$=1) the transition time is explicitly related to the current according to

$$\tau_{app} = \frac{-1}{\pi^2}\ln\left[\frac{\pi^2}{8}\left(1 - \frac{1}{|i|}\right)\right], \quad (29)$$

which is valid for a relatively short transition time or equivalently for small applied currents, just above the diffusion-limiting current. For large applied currents (or short transition times) the similarity solution, i.e. eq. (16), will result in a more accurate prediction of the ion concentration, $c$. Therefore, in this case we can use eq. (16) to determine the ion concentration at the inner-outer region interface. We now substitute either $x$=1 for a positive current (or $x$=0 for a negative current) in eq. (16) and use the relations $\text{ierfc}(0) = 1/\sqrt{\pi}$ and $\text{ierfc}[1/(2\cdot\sqrt{\tau})] = 0$ when $\tau \to 0$ to obtain the well-known Sand equation,[15]

$$\tau_{Sand} = \frac{\pi}{16 \cdot i^2}, \tag{30}$$

for the transition time.

Next we combine the approximate solutions for the transition time for small and large applied currents, i.e. eqs. (29) and (30), respectively. As a result we obtain a single analytical equation describing the whole current domain, ranging from small to large applied currents, according to

$$\tau_{tr} = (1-h) \cdot \tau_{Sand} + h \cdot \tau_{app}, \tag{31}$$

where $h(i) = \exp(-(i-1)^2/\sqrt{2})$, a function which equals approximately unity at currents in the vicinity of the diffusion-limiting current, i.e. $i \sim 1$, and becomes zero at large currents, while having a value of ~0.5 when eqs. (29) and (30) coincide.

The analytical solution for the transition time according to eq. (31) is valid for applied currents above the diffusion-limiting current. For an applied current exactly equal to the diffusion-limiting current we find an interesting phenomenon. Namely, in the case of $i=1$ the transition time will become infinite as eq. (15) tends asymptotically to the steady-state, and thus it takes infinite time to reach a zero ion concentration. Interestingly, if we substitute the diffusion-limiting current, i.e. $i=1$, into eq. (25) and consider the long term behavior, i.e. $\tau \gg 1$, we obtain

$$\Delta\phi_{outer}(i=1) \approx \pi^2\tau + 2 \cdot \ln\left(\frac{\pi}{2}\right) \tag{32}$$

for the potential drop across the outer region. Eq. (32) does not show an asymptotic behavior, but shows that the potential increases linearly with time, see ref. 46 (p. 212).

## 3    Results and Discussion

In this section we present results for the cell potential as function of time for various kinetic constants and values of the imposed currents for both the thin-DL limit and the full PNP model as described above. The results for the thin-DL limit are easily obtained using a simple numerical routine to solve eqs. (6), (13)-(15), (17) and (18) simultaneously. The GC- and H-limit can be solved directly as they are described by eqs. (26) and (27), respectively. The full PNP model results are obtained by finite element discretization using the commercially available COMSOL software package. We have used a numerical grid with a variable spacing, namely, near the electrodes the spacing was at least one tenth of $\epsilon$, while away from the edges the spacing was never larger than $\Delta x$=0.01. In all cases the initial cat- and anion fluxes are set to zero, such that the system is at rest at $\tau$=0. From $\tau$=0 onward we use a simple time stepping routine with a relative tolerance on the time steps of $10^{-3}$ in combination with a direct solving method to compute the time-dependent behavior of the electrochemical cell on applying a constant current.

To show the conditions for which the thin-DL limit is appropriate we first present a comparison between this limit and the full model results. Next, we present results for the GC- and H-limit and compare these with the results for various values of the Stern layer thickness. Furthermore, we show results for asymptotic behavior at the transition time for the thin-DL limit when the current is increased to values above the diffusion-limiting current. Finally, we will present full model results for currents above the diffusion limit, which show that for finite values of $\epsilon$ this asymptotic behavior will not occur.

### 3.1 Thin-DL limit

First we show results for one set of kinetic constants while using two values of the applied current, namely, $i$=0.25 and 0.95, while $k_R$=$j_O$=10 and $\delta$=1. These numbers are chosen such that $i$=0.25 represents a relatively small current, i.e. close to the open-circuit condition, whereas $i$=0.95 is close to the diffusion-limiting current of $i$=1. Due to the choice of equal values for $k_R$ and $j_O$ at both electrodes an initial DL on either electrode is absent, i.e. at $\tau$=0 the concentration, $c$, equals unity across the complete electrolyte phase. As a result the open-cell potential equals zero. In Fig. 2 we show results for the values of the parameters discussed above, and for values of $\epsilon$ ranging from 0 to $10^{-2}$.

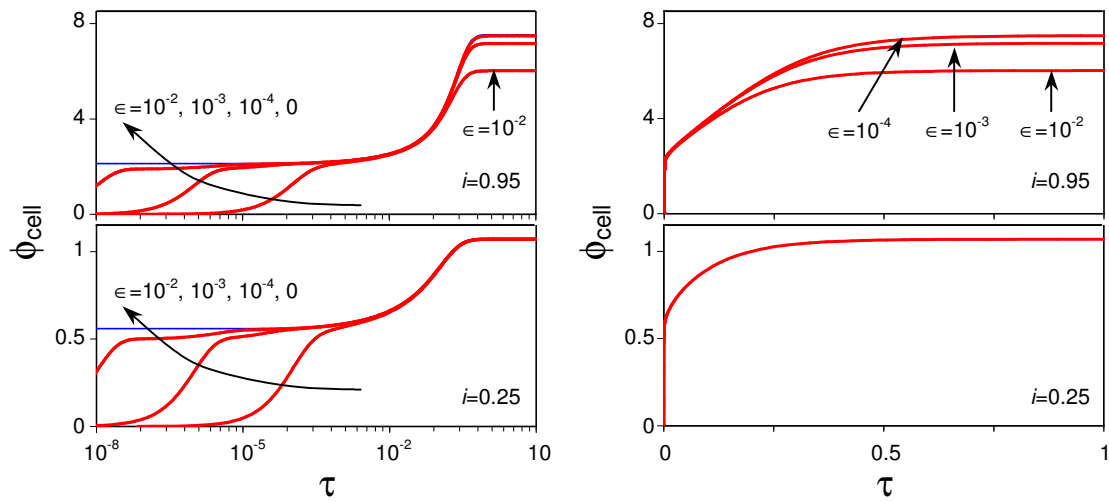

Fig. 2. (Color online) Cell potential as function of time for $i$=0.25 and 0.95, $\epsilon$=0..$10^{-2}$, $k_R$=$j_O$=10 and $\delta$=1 (The arrows indicate increasing values of $\epsilon$).

Interestingly, in case of the thin-DL limit the applied current, $i$, is purely Faradaic at the reaction plane during the complete transient response of the cell, while for any non-zero value of $\epsilon$ initially this current is of a purely capacitive nature due to the formation of the DLs, which is represented by the Maxwell current. The Maxwell current will result in an increasing electrical field at the electrodes and thus subsequently in an increasing Faradaic current via the increasing potential drop across the Stern layer. As a result the electrical field strength will increase until the Faradaic current reaches the value of the applied current. Consequently, for $\epsilon$>0 we have initially a zero cell potential which increases due to the Maxwell current, while for the thin-DL limit the cell potential is immediately non-zero and constant until the ions start to redistribute across the outer region, which is at the diffusion time-scale. The initial rise-time for the cell potential in case of $\epsilon$>0 is approximately $\tau$=$\epsilon$ as shown in Fig. 2. This is in line with calculations for cells without Faradaic reactions at their electrodes, namely, for these cells it was shown that the relaxation time-scale, i.e. $\tau_r$=$\tau/\epsilon$, is the characteristic time-scale.[18] Consequently, the rise-time will decrease for decreasing values of $\epsilon$, such that the thin-DL limit becomes more accurate when $\epsilon$ decreases. As a result, the model based on the thin-DL limit, which consists of a small set of algebraic equations for each time, is exactly equivalent to the full PNP-gFBV theory in the limit of $\epsilon \to 0$.

Considering the evolution of the cell potential (Fig. 2), we observe that from its initial plateau, the cell potential further increases while the ions redistribute across the cell at the longer diffusion time-scale. At and beyond this time-scale a perfect match is observed between the thin-DL limit and full PNP model calculations when sufficiently small values of $\epsilon$ are used. For $i$=0.95, the steady-state potential decreases for increasing values of $\epsilon$. This corresponds to results presented in earlier work[31] where it was shown that the steady-state cell potential at low currents coincides for all values of $\epsilon$, while it diverges for relatively large values of $\epsilon$ when approaching the diffusion-limiting current, i.e. for currents approaching $i=\pm 1$.[31] As a result, the accuracy of the thin-DL limit does not only depend on the value of $\epsilon$ but also on the value of the applied current. However, here we find that $\epsilon$ equal to $10^{-4}$ results in a near perfect match with the thin-DL limit at the diffusion time-scale for currents up to $i$=0.95.

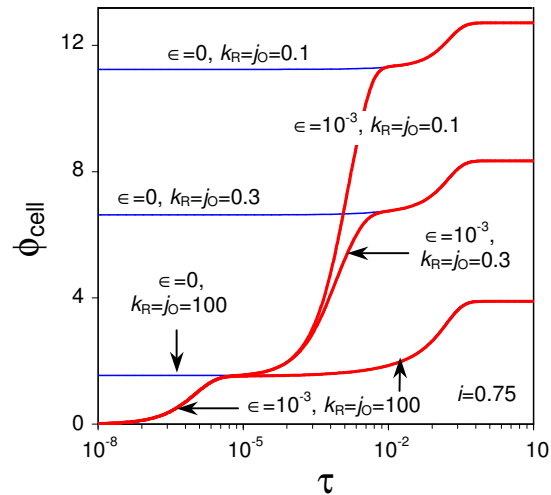

Fig. 3. (Color online) Cell potential as function of time for $i$=0.75, $\epsilon$=0 and $10^{-3}$, $k_R=j_O$=0.1..100 and $\delta$=1.

Next, we show results for three values of the kinetic constants, namely, $k_R$ and $j_O$ equal to 0.1, 0.3 and 100, see Fig. 3. These constants represent relatively slow ($k_R=j_O$=0.1 and 0.3) and fast ($k_R=j_O$=100) reaction kinetics at the electrodes. Again the results for both models, as shown in Fig. 3, coincide for the times at and beyond the diffusion time-scale. The results for $\epsilon=10^{-3}$ (red lines) show that for low values of the kinetic constants an additional plateau value for the cell potential will appear earlier and below the plateaus as discussed above. The presence of these three plateau values can be explained as follows. The applied current will first result in an increase of the potential drop across the outer region,[59] as indicated in Fig. 4a, leading to the first plateau value. Next, the potential drop across the outer region results in an ion flux across this region and consequently in the formation of the DLs, indicated by the increase of the potential drop across the Stern layer (Fig. 4b), leading to the second plateau value. Finally, the potential drop across the Stern layer results in the increase of Faradaic current at the reaction planes (Fig. 4c) and consequently in a redistribution of ions across the outer region, which leads to the third and final plateau value. For high kinetic constants the first and second plateau value are indistinguishable as the potential drop across the electrochemical double layer remains small. For fast reaction kinetics this potential drop can remain small since only a small variation in the potential drop across the Stern layer is required for the Faradaic current to achieve the applied current.

Furthermore, we can observe in Fig. 4c that the Faradaic current at $x$=1 starts to increase before the Faradaic current at $x$=0, with the result that when the steady-state is reached we have a slight decrease of the total number of cations in the cell, i.e. the electrolyte contains less cations than anions, a situation different from classical treatments where electro-neutrality is assumed a priori in the entire cell.

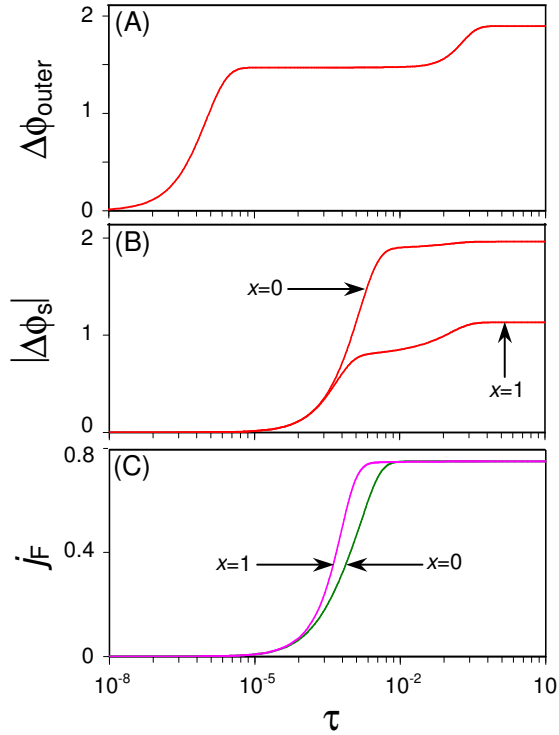

*Fig. 4. (Color online) Time evolution of (A) the potential drop across the outer region, (B) the potential drop across the Stern layer, and (C) the Faradaic current; i=0.75, $k_R$=$j_O$=0.3, $\delta$=1.*

*3.2 GC- and H-limit*

The analytical expressions for the GC- and H-limit, i.e. eqs. (26) and (27), are based on the first order approximation of the diffusion equation, and are thus only a good approximation if the ion concentration is nearly linear across the outer region. In Fig. 5a we show the ion concentration distribution in the outer region for different times for the thin-DL limit below the diffusion-limiting current. Fig. 5a shows that initially the applied current influences the ion concentration near the edges of the cell only, while the concentration profile becomes linear when the steady-state is approached. We are already close to linear behavior at $\tau$=0.075, which indicates that the analytical eqs. (26) and (27) are a good approximation for the GC- and H-limit, respectively. For currents above the diffusion-limiting current we do not have a linear behavior of the ion concentration as indicated in Fig. 5b since we reach the transition time at $\tau$~7.85x10$^{-3}$, as we will discuss in section 3.3. In Fig. 6 we present results for eqs. (26) and (27) and for the thin-DL limit with varying values for the Stern layer thickness.

In Fig. 6a we present results for an electrolytic cell, i.e. the cell has a zero open cell potential and can thus not auto-generate current, which is the result of the equal kinetic parameters ($k_R=j_O=10$) at both electrodes. The results show that initially the GC- and H-limit coincide, while at the steady-state there is a small difference in cell potential between both these limits. The reason for the initial overlap of both curves is the relatively high value of the kinetic constants, which result in a relatively small potential drop across either the diffuse (GC-limit) or Stern (H-limit) layer. At longer times the ions will redistribute across the outer region, such that the potential drop across the Stern or DL will increase, which results in a difference in cell potential for both limits. This can also be seen mathematically from the discussion below eqs. (26) and (27) in the theory section. There it was shown that the second term of eq. (26) vanishes for large values of the kinetic constants, while the second and third term of eq. (27) vanish as function of time for the same condition. Thus, it is possible that at short times the two limits give the same results, while a difference only develops when the steady-state is reached. The inset of Fig. 6a shows the steady-state cell potential as function of the Stern layer thickness. It shows that for $\delta=10^{-2}$ and $\delta=10^4$ the cell potential for the GC- and H-limit is closely approached, while for a more realistic value of $\delta$ in the order of unity neither limit is exact.

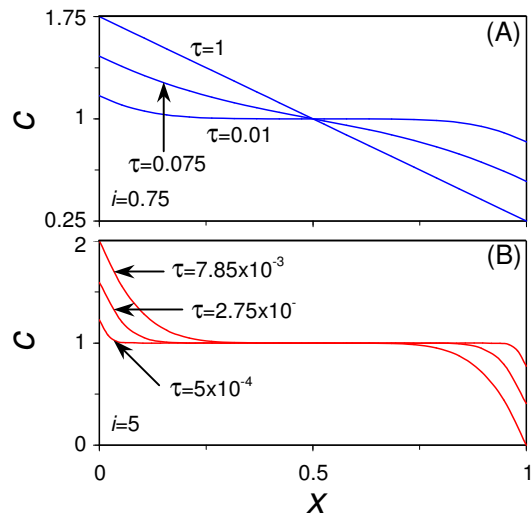

Fig. 5. (Color online) Concentration profiles in the outer region for the thin-DL limit for (A) $i=0.75$, $\tau=0.01$, 0.075 and 1, and for (B) $i=5$, $\tau=5\times10^{-4}$, $2.75\times10^{-3}$ and $7.85\times10^{-3}$.

In Fig. 6b we present results for a galvanic cell, i.e. the cell has a non-zero open cell potential and can thus auto-generate current. The kinetic constants for this cell are $k_{R,A}=300$, $j_{O,A}=1$, $k_{R,C}=10$, $j_{O,C}=8$, such that the open cell potential equals $\phi_0=5.48$. The time-dependent behavior of the cell potential for this case is very different compared to the previously described electrolytic cell. This is due to the fact that the applied current ($i=0.95$) is large compared to the oxidation rate constant at the anode ($j_{O,A}=1$), which results in a large potential drop across the DL there. The increase in cell potential in this case follows mathematically from the second term of eq. (26), as discussed in the theory section. The inset of Fig. 6b shows that the transition from the GC- to the H-limit for increasing values of the Stern layer thickness differs from the behavior for the electrolytic cell plotted in the inset of Fig. 6a. Here, in Fig. 6b, the curve for increasing value of the Stern layer thickness shows a non-monotonic behavior, which can even break the 'limiting' value of the H-limit, as already mentioned in ref. 23. Still at sufficiently small or large values of the Stern layer thickness the steady-state cell potential will converge to the GC- and H-limit, respectively.

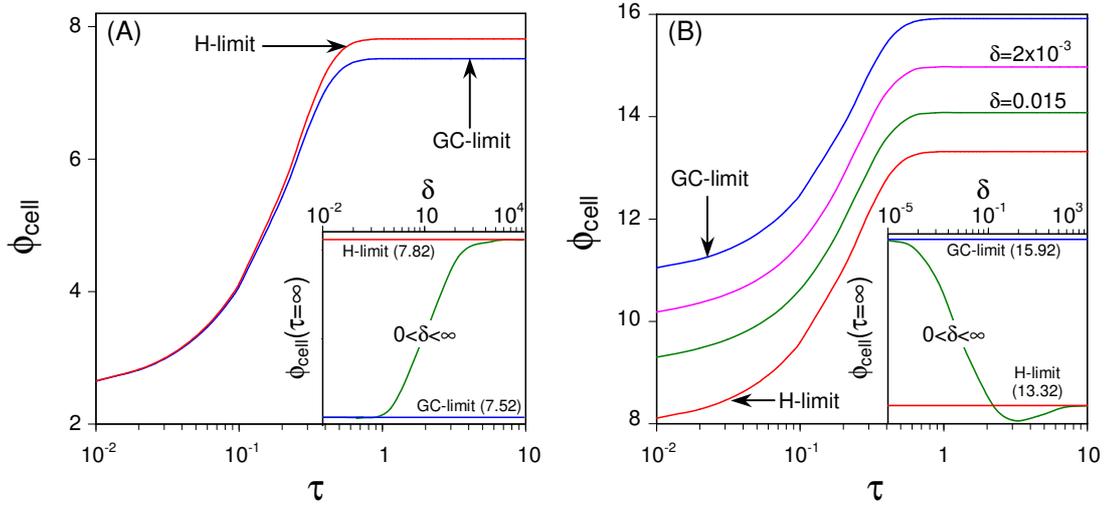

*Fig. 6. (Color online) Cell potential as function of time for the thin-DL limit, for arbitrary $\delta$ and in the GC- and H-limit; i=0.95; (A) An electrolytic cell with identical electrodes, $k_R=j_O=10$; (B) A galvanic cell with different electrodes and a non-zero open circuit voltage, $k_{R,A}=300$, $j_{O,A}=1$, $k_{R,C}=10$, $j_{O,C}=8$; Insets: steady-state current as function of the Stern layer thickness.*

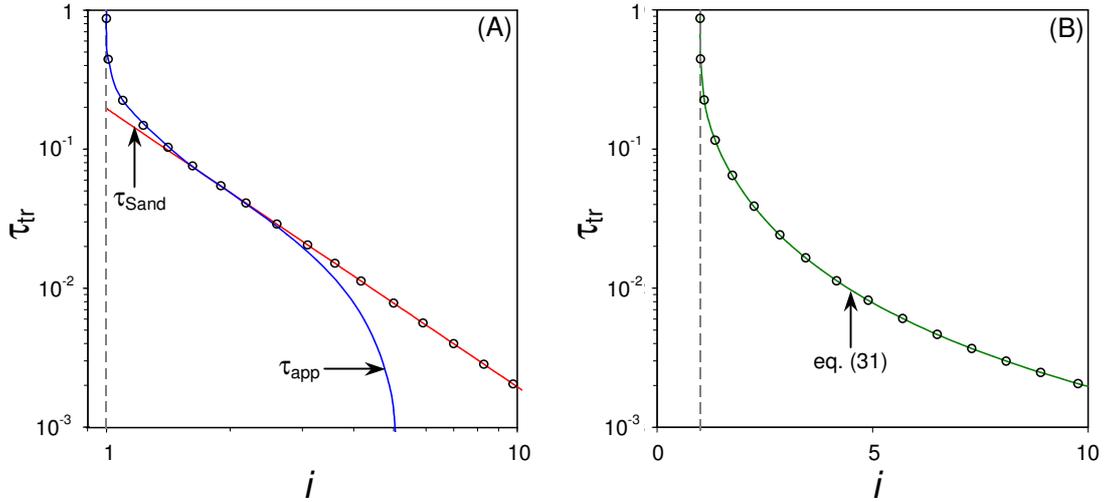

*Fig. 7. (Color online) Transition time as function of the applied current according to (A) the two approximate solutions $\tau_{Sand}$ and $\tau_{app}$, and (B) the combined solution according to eq. (31), the open circles in both panels indicate the exact solution according to eq. (28).*

### 3.3 Transition time

Next we show results for a cell with an applied current above the diffusion-limiting current. If we consider the thin-DL limit the cell potential at these currents will diverge if the transition time is reached. In the previous section we showed that the implicit relation given by eq. (28) can be used to determine the exact solution for the transition time. The explicit relation for the transition time, eq. (31), was obtained by combining the solutions using a first order approximation for the diffusion equation, $\tau_{app}$, and the Sand equation, $\tau_{Sand}$. The results for both $\tau_{app}$ and $\tau_{Sand}$, presented in Fig. 7a, coincide with the results for eq. (28) in the low and high applied current regime, respectively, while they intersect at *i* equal to ~2. The reason for the perfect match of $\tau_{app}$ and $\tau_{Sand}$ in their corresponding regime can be explained as follows. In the low current regime the ion concentration profile across the outer region is nearly linear for longer times (Fig. 5a), a profile that is well described by the first order approximation of eq. (15). As a result $\tau_{app}$ coincides with the exact solution given by eq. (28) in the low current regime. In the high current regime we can distinguish two separate diffusion layers near the electrode (Fig. 5b), such that the similarity solution given by eq. (16) results in an accurate description of the ion concentration. Consequently, the Sand equation, $\tau_{Sand}$, accurately describes the transition time in the high current regime. (Note that in Fig. 5b $\tau_{tr}$ equals ~7x85·10$^{-3}$.) In Fig. 7b we show the results for the combined equation, eq. (31). These results show that eqs. (28) and (31) coincide in the low as well as in the high current regime. As a result we can use eq. (31) as very accurate analytical equation in place of the exact, implicit relation, eq. (28).

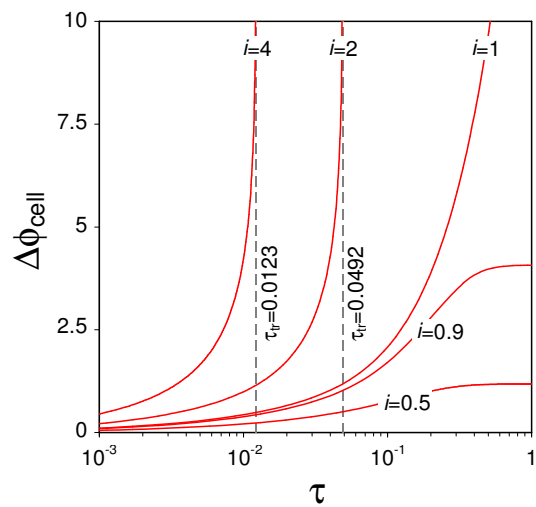

*Fig. 8. (Color online) Increase in cell potential as function of time for currents below and above the diffusion-limiting current (full lines) including the corresponding transition time (dashed lines) for the thin-DL limit; i=0.5, 0.9, 1, 2, 4, $k_R=j_O$=10, and $\delta$=1.*

In Fig. 8 we show results for the increase in cell potential in case of the thin-DL limit as function of time for various values of the applied current. The increase in cell potential is defined as the difference between the actual cell potential and the initial plateau value, i.e. $\Delta\phi_{cell} = \phi_{cell}(\tau) - \phi_{cell}(\tau = 0)$. Again we assume fast reaction kinetics, i.e. $k_R=j_O$=10, and set $\delta$ equal to 1. We can see from Fig. 8 that for *i*=0.5 and *i*=0.9 the cell potential reaches a plateau value at the steady-state just as in Fig. 2. However, for an applied current equal to the diffusion-limiting current, i.e. *i*=1, the cell potential will increase unbounded and it will not reach a vertical asymptote as will occur for currents above the diffusion-limiting current. This behavior at the diffusion-limiting current is due to fact that at the diffusion-limiting current the concentration of reactive ions will only reach zero concentration for $\tau$ tending to infinity as explained in the theory section. At higher currents the reactive ions will reach zero concentration in finite time as indicated in Fig. 8 for *i*=2 and *i*=4 by the dashed lines.

In previous work[31,32,57] it was shown that the steady-state current can break the diffusion-limiting current if a finite ratio of Debye length, $\lambda_D$, to electrode spacing, $L$, is assumed, i.e. $\epsilon > 0$. In Fig. 9 we present results for the cell potential as function of time for a system operating at $i=2$ for the thin-DL limit and $\epsilon$ equal $10^{-3}$ and $10^{-2}$. These results show that a system with a finite value for $\epsilon$ can break the vertical asymptote at the transition time due to the expansion of the DL at the cathode ($x=1$) into the bulk region.[31,32,57]

In Fig. 10a we show the space-charge density profiles across the system for various time steps. Fig. 10a shows that below the transition time the thickness of the DL is smaller than the theoretical thickness of $\epsilon^{2/3}$.[31,32] However, for longer time the space-charge region at the cathode expands into the bulk solution, such that a charge-free outer region can no longer be assumed. The DL at the anode, however, remains within its theoretical boundary, i.e. does not expand to macroscopic dimensions.

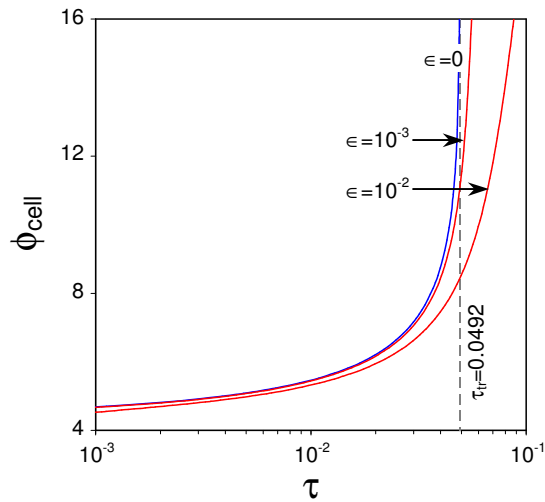

*Fig. 9. (Color online) Cell potential as function of time for a current above the diffusion-limiting current for both the thin-DL limit and full model, dashed is the transition time; $\epsilon=0$, $10^{-3}$, $10^{-2}$, $i=2$, $k_R=j_O=10$, and $\delta=1$.*

To summarize, we have presented results for the evolution in time of the cell potential of electrochemical cells for various parameter settings. For all the parameters that we have investigated we have obtained a good fit between the thin-DL limit and the full PNP model for sufficiently small values of $\epsilon$, except for a very brief initial period. These initial deviations can be contributed to the influence of the Maxwell current, which is included in the full PNP model but neglected in the simplified model. Because the Maxwell current effectively vanishes beyond a characteristic time of $\tau=\epsilon$, the simplified model becomes more accurate at short times when $\epsilon$ decreases (i.e., for larger system dimensions relative to the Debye length). Furthermore, we have shown that during the start-up of an electrochemical we can distinguish three plateau values for the cell potential, namely, a first plateau value due to an increasing potential drop across the bulk region, a second plateau due to the formation of the DLs and finally a third plateau value due to the redistribution of ions across the bulk region. Next, we have shown the influence of the Stern layer thickness on the cell potential by considering the GC- and H-limit. We have shown that for an electrolytic cell with high kinetic constants the influence of the Stern layer thickness on the evolution of the cell potential is small. For a galvanic cell with the value of one of the kinetic constant close to the applied current there is a clear distinction in evolution of the cell potential between the GC- and H-limit, showing that in this case the influence of the Stern layer thickness is significant. Finally, we have shown that the cell potential for the thin-DL limit reaches a vertical asymptote at the transition time if the applied current is larger than the diffusion-limiting current, a limit than can be broken due to the expansion of the DL near the cathode for models where $\epsilon > 0$.

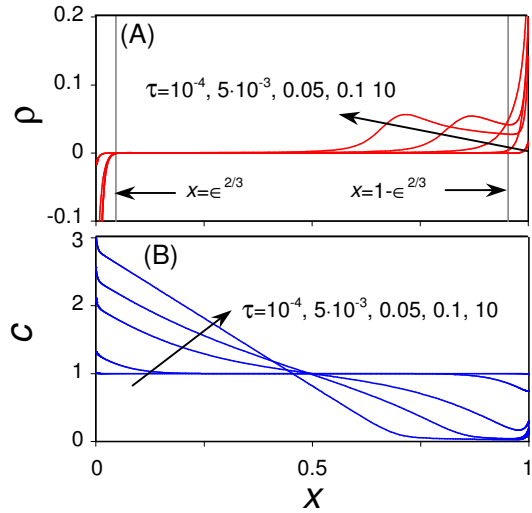

Fig. 10. (Color online) Time evolution of (A) the space-charge density distribution and (B) the concentration profiles, vertical lines in (A) indicate the theoretical 'inner-outer region' interfaces; $\epsilon=10^{-2}$, $i=2$, $k_R=j_O=10$, and $\delta=1$ (The arrows indicate increasing values of $\tau$).

## 4 Conclusions

We have presented a mathematically simplified model for the transient (dynamic, time-dependent) voltage of an electrochemical cell in response to a current step, including diffuse charge effects, for a one-dimensional and planar geometry in the limit of a negligibly thin Debye length compared to the electrode spacing. The simplified model couples a diffusion equation for the interior of the cell to analytic boundary equations that describe the diffuse layers (space-charge region near the electrodes) and the Faradaic reaction kinetics. As a result, the simplified model significantly reduces the numerical complexity of the full model, which is based on the generalized Frumkin-Butler-Volmer equation for the electrochemical reaction at the electrodes and the Poisson-Nernst-Planck equation for the transport of ions throughout the electrolyte. Further, we have presented analytical equations for the evolution of the cell potential in time based on a first order approximation of the diffusion equation and the additional assumption of either a zero or infinite reaction plane to electrode spacing compared to the Debye length. The first order approximation for the diffusion equation was also used to extend Sand's equation to the domain of large transition time (or equivalently currents just above the diffusion-limiting current).

We have shown that for applied currents below the diffusion-limiting current the simplified model is accurate at and beyond the diffusion time-scale when the Debye length is small compared to the electrode spacing. For super-limiting currents the simplified model is valid up to the transition time at which the cell voltage blows up due to ion depletion in the electro-neutral bulk electrolyte. In the full model we can have super-limiting currents beyond the transition time as the diffuse layer expands to a non-equilibrium structure resulting in charging of the bulk electrolyte. Consequently, the simplified model can be very beneficial in modeling the transient response of electrochemical cells up to the steady-state for currents below the diffusion-limiting current, or up to the transition time for super-limiting currents, if the Debye length is small compared to the feature size of the cell, an assumption which is generally valid since the Debye length is typically in the order of 1-100 nm.

In conclusion, taking into account diffuse charge effects, and thus not assuming electro-neutrality in the entire cell, leads to a more insightful and comprehensive model, while the mathematical complexity of the model does not increase significantly compared to the classical models based on electro-neutrality in the entire cell.

**Acknowledgments**

This research was carried out under project number MC3.05236 in the framework of the Research Program of the Materials innovation institute M2i (www.m2i.nl). MZB also acknowledges support from U.S. National Science Foundation under contracts DMS-0842504 and DMS-0855011.